\documentclass[aps,showpacs,twocolumn,floats,prl]{revtex4}
\usepackage{graphicx}
\usepackage{amsmath}
\usepackage{psfrag}

\newcommand {\e} {\varepsilon}                  % shortcuts for some
\newcommand{\be}{\begin{equation}}              % redefines \begin{equation}
\newcommand{\ee}[1]{\label{#1} \end{equation}}  % redefines \end{equation}

\newcommand{\bee}{\begin{eqnarray}}             % redefines \begin{eqnarray}
\newcommand{\eee}{\end{eqnarray}}               % redefines \end{eqnarray}
\def\reff#1{(\ref{#1})}

\begin{document}

\title{Mixing-induced global modes in open active flow
}
\author{Arthur V. Straube\footnote{E-mail: arthur.straube@gmail.com \\[0.5mm] {\mbox{} Paper published in Phys. Rev. Lett {\bf{99}}, 184503 (2007)}} and Arkady~Pikovsky}
\affiliation{Department of Physics, University of Potsdam, Am
Neuen Palais 10, PF 601553, D-14415, Potsdam, Germany}
%\homepage{www.agnld.uni-potsdam.de}
\date{\today}
\begin{abstract}
We describe how local mixing transforms a convectively unstable
active field in an open flow into absolutely unstable. Presenting
the mixing region as one with a locally enhanced effective
diffusion allows us to find the linear transition point to an
unstable global mode analytically. We derive the critical exponent
that characterizes weakly nonlinear regimes beyond the instability
threshold and compare it with numerical simulations of a full
two-dimensional flow problem.
 \end{abstract}

%\pacs{05.45.-a}%Nonlinear dynamics and nonlinear dynamical systems
%\pacs{47.70.-n}%Reactive, radiative or nonequilibrium flows
%\pacs{47.52.+j}%Chaos
%\pacs{47.54.-r}%pattern selection, pattern formation
\maketitle

In many natural and laboratory flows active chemical and
biological processes may occur. Examples include chemical
reactions in minimixers, plankton growth in ocean, see e.g.
\cite{Tel_etal-05} and references therein. Quite often this
activity occurs in an open rather than in a closed geometry. Here
the main issue is whether the throughflow is stronger or weaker
than the activity. One has to compare the velocity of the
throughflow with the velocity of the activity spreading due to
diffusion. If the throughflow is stronger, the activity is ``blown
away'' like a candle flame in a strong wind, in the opposite case
a sustained activity can be observed \cite{refs24}. This simple
picture is valid, however, only for homogeneous media. Often
additional vortexes are superimposed on a constant throughflow,
due to, e.g., wakes behind islands in ocean currents or mixing
enforced by revolving fan blades in laboratory experiments. We
want to study under which conditions such an additional kinematic
mixing in a strong open flow can lead to a transition to a
sustained activity, and to characterize this transition
quantitatively.

Our main model is a reaction-advection-diffusion equation for the
dimensionless concentration of an active scalar field
$u(\mathbf{r},t)$
\be
%\frac{\partial u}{\partial t}
u_t +(\mathbf{V} +\mathbf{W}(\mathbf{r},t))\cdot \nabla
u=D_0\nabla^2 u+au(1-u^p)\;.
\ee{eq1}
Here $\mathbf{V}=(V,0,0)$ is a constant throughflow in
$x$-direction, $D_0$ is a molecular diffusion of the scalar field.
Activity is assumed to be of the simplest form: a linear growth
with rate $a$ with a saturation at $u=1$. The nonlinearity index
$p$ is typically integer (1 or 2) for chemical reactions, while
for biological populations a wide range of values of $p$ has been
recently reported~\cite{Sibly-etal-05}. Mixing is described by a
spatially localized incompressible velocity field
$\mathbf{W}(\mathbf{r},t)$, its intensity is denoted as $W$. Note
that in the absence of fluid flow Eq.~\reff{eq1} is reduced to the
famous Kolmogorov-Petrovsky-Piskunov-Fisher (KPPF) model of an
active medium with diffusion (see, e.g., \cite{Murray-93} for
original references, analysis, and applications of KPPF), while
for $a=0$ Eq.~\reff{eq1} describes a linear evolution of a passive
scalar in a flow. Model \reff{eq1} can be used for the description
of biological activity, where $u$ is, e.g., concentration of a
growing plankton, advected by oceanic currents; for a possible
laboratory realization see recent experiments  with an
autocatalytic reaction
%
% between iodate and arsenous acid
%
in a Hele-Shaw cell with a
throughflow~\cite{Leconte-Martin-Rakotomalala-Salin-03}.

In the absence of flow, the diffusion causes the active state to
spread forming eventually a front with velocity
$V_f=2\sqrt{aD_0}$~\cite{Saarloos-03}. Thus, for vanishing mixing
$W=0$, the activity is blown away provided $V>V_f$. For this
parameter range the instability in Eq.~\reff{eq1} is convective
and in the absence of external sources, no activity is observed in
the medium. A nontrivial state is, however, possible if there is a
localized source of the field $u$: then a growing tail stretches
from this source in the downstream direction, where it eventually
saturates at $u=1$. The linearized problem with a point
($\delta$-function) source of intensity $\e$ can be readily
solved, yielding
\begin{align}
 u(x)&=\e(\tilde V)^{-1}\exp[xV/2D_0]\exp(-|x|\tilde V/2D_0)\;,
\label{eq01}\\
 u(\mathbf{r})&=\e(2\pi D_0)^{-1}\exp[xV/2D_0]K_0(|\mathbf{r}|\tilde V/2D_0)
\label{eq02}
\end{align}
in  one- and two-dimensional setups
% and
%
%
%in two dimensions
(where $K_0$ is the modified Bessel function,
$\tilde V=(V^2-V_f^2)^{1/2}$).
%$\tilde V=\sqrt{V^2-V_f^2}$.
Note that in both solutions
$u\sim\exp(\mu_{\mp} x)$ as $x\to\pm\infty$, where
$\mu_{\pm}=(2D_0)^{-1}(V\pm\tilde V)$.

\begin{figure}%[!htb]
\centering
\includegraphics[width=0.3\textwidth]{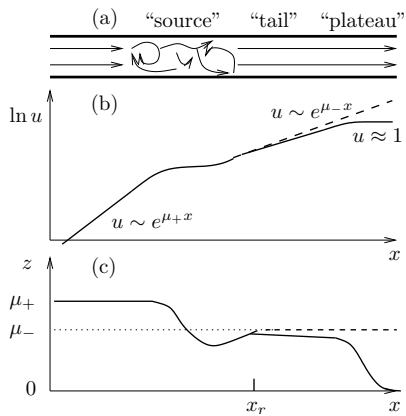}
\caption{(a): Quasi-one-dimensional open flow with a localized
mixing zone. Panels (b) and (c) illustrate the construction of the
nonlinear global mode, they show qualitative profiles $u(x)$ and
$z(x)=u^{-1}\frac{du}{dx}$ in the linear approximation at the
criticality (dashed line) and in nonlinear regime slightly beyond
criticality (full line).  In the latter case the profile is nearly
linear for $x<x_r$ but deviates due to nonlinear terms for
$x>x_r$, see discussion of Eq.~(\ref{eq10}). Regions ``source,''
``tail,'' and ``plateau'' are explained in the text.} \label{fig1}
\end{figure}

In this letter we demonstrate, that beyond some critical intensity
$W_{cr}$, a localized mixing $\mathbf{W}(\mathbf{r},t)$ turns the
convective instability locally into the absolute one, which
results in a stationary (in statistical sense) profile of $u$ (see
Fig.~\ref{fig1} for a sketch of the profile and Fig.~\ref{fig5}
below for a numerical example). Beyond criticality
$W>W_{cr}$, the mixing region acts as an effective source of the
 field, in Fig.~\ref{fig1} this region is denoted as a
``source.'' A ``tail'' where the field grows exponentially as in
(\ref{eq01}), (\ref{eq02}) extends downstream of the source.
Further downstream stretches the ``plateau'' domain where $u=1$.
Our main quantitative result, obtained by matching solutions in
these three domains, is the critical exponent $\beta$ relating the
intensity of the effective source $\e_{eff}$ with the mixing
intensity: $\e_{eff}\sim (W-W_{cr})^\beta$.

To develop the theory we use the
concept of global modes~\cite{Chomaz-Huerre-Redekopp-88%,Couairon-Chomaz-97,Tobias-Proktor-Knobloch-98
}. In this concept a self-sustained non-advected pattern arises
due to inhomogeneities of the system. Typically, one considers
inhomogeneities of the growth rate $a$: if $a=a(\mathbf{r})$ has a
hump where locally the front velocity is large $V_{f\;loc}>V$,
then a global mode appears, located at this hump and downstream.
In this paper we are interested in another, mixing-based mechanism
of a global mode appearance. It can be easily understood if the
concept of effective diffusion (see, e.g.,~\cite{Dimotakis-05}) is
used to describe the mixing term in \reff{eq1}. In this approach,
we phenomenologically introduce effective diffusivity
$D(\mathbf{r})=D_0+D_{mix}(\mathbf{r})$ that accounts for the
coarse grained mixing dynamics, and write instead of \reff{eq1} an
equation with a non-homogeneous diffusion
\be
%\frac{\partial u}{\partial t}
u_t +\mathbf{V}\cdot \nabla u=\nabla [D(\mathbf{r})\nabla
u]+au(1-u^p)\;. \ee{eq3}
A hump of diffusivity $D(\mathbf{r})$ leads to an increase of
the local front velocity $V_f$, and one expects that when the
front propagation prevails over the throughflow, a stationary
global mode can appear, producing a mixing-induced sustained
structure. We focus on a geometry shown in
Fig.~\ref{fig1}(a): in a constant open flow there is a localized
region of strong mixing, which, as we will see below, is not
necessarily chaotic or turbulent. The theory below will be
developed for a one-dimensional case, which is relevant, e.g., for
flows in a micropipe; the results will be supported by numerical
simulations of two-dimensional flows. We restrict ourselves to
this case because of computational simplicity, and also because
two-dimensional flows are relevant for many geophysical and
laboratory (especially in microfluidics) experimental situations.

We start with a linear analysis of a one-dimensional situation, described
by the linearized at $u=0$ Eq.~\reff{eq3}:
\be \frac{\partial u}{\partial t}+ V\frac{\partial u}{\partial
x}=\frac{\partial}{\partial x} \left[D(x)\frac{\partial
u}{\partial x}\right]+au\;. \ee{eq4}
We look for an exponentially growing in time solution and with an
ansatz $u(x,t)\sim \exp[\lambda t+\int^x z(\xi) d\xi]$ obtain
\be \frac{dz}{dx}=-z^2+\frac{V-\frac{d
D(x)}{dx}}{D(x)}z-\frac{a-\lambda}{D(x)}\;. \ee{eq5}
As $|x|\to \infty$ we have a homogeneous medium with $D=D_0$, here
the solution should tend to values
$z^0_{\pm}=(2D_0)^{-1}(V\pm\sqrt{V^2-4(a-\lambda)D_0})$ at which
the r.h.s. of \reff{eq5} vanishes. More precisely, as
$x\to-\infty$ we have $z\to z^0_+$ and as $x\to\infty$ we have
$z\to z^0_-$. With these two boundary conditions one easily finds
the solution of \reff{eq5} numerically, matching at $z=0$
integrations starting at large $|x|$ from the values $z^0_\pm$. In
a particular analytically solvable case of a piecewise-constant
diffusivity: $D=D_0$ for $|x|>l$ and $D=D_1>D_0$ for $|x|<l$, one
can perform the integration analytically and obtain the equation
for the growth rate $\lambda$:
\be
l=\frac{2D_1}{\sqrt{4(a-\lambda)D_1-V^2}}\arctan\sqrt{\frac{V^2-4(a-\lambda)
D_0 } {4(a-\lambda)D_1-V^2 } }\;. \ee{eq6}
The value $\lambda=0$ corresponds to the onset of global
instability, in this case \reff{eq6} gives the relation between
the critical values $l_{cr}$ and $D_{1cr}$:
\be
l_{cr}=\frac{2D_{1cr}}{\sqrt{4aD_{1cr}-V^2}}\arctan\sqrt{\frac{V^2-V_f^2
} {4aD_{1cr}-V^2 } }\;. \ee{eq7}
From \reff{eq7} it follows that $D_{1cr}\to\infty$ as $l_{cr}\to
l_{min}=\tilde V /(2a)$. In other words, there exists a minimal
size of the mixing region, so that for $l<l_{min}$ even a very
strong mixing, with a very large effective diffusion, cannot
create a global mode (the same is true for a smooth profile of
$D(x)$; note also that the size of the mixing region is not
limited from above). This is in contrast to the situation when the
global mode is induced by a local hump of the growth rate $a$
(cf.~\cite{Dahmen-Nelson-Shnerb-00}): here one can obtain
instability even when $a(x)$ is highly localized (a
delta-function), a global mode then looks as in \reff{eq01},
\reff{eq02}. A similar analysis performed for a two-dimensional
inhomogeneous version of Eq.~\reff{eq4} also yields a minimal
radius of a diffusive spot that can lead to instability.

Next we discuss the linear stability not in the framework of the
effective diffusion model~\reff{eq3}, but in the full
reaction-advection-diffusion problem as in Eq.~\reff{eq1}. After
the linearization we arrive at a linear stability problem
\be
%\frac{\partial u}{\partial t}
u_t
+(\mathbf{V}
+\mathbf{W}(\mathbf{r},t))\cdot \nabla u=D_0\nabla^2 u+au\;,
\ee{eq801}
which is nonstationary if the velocity field $\mathbf{W}$ is time
dependent. Then the proper way to determine the stability is to
calculate the largest Lyapunov exponent (LE) $\Lambda=\langle
\frac{d}{dt}\ln||u||\rangle$. This can be done numerically, as
described in ref.~\cite{Pikovsky-Popovych-03}. Noteworthy, in this
consideration we are not restricted to a deterministic flow, as
the LE can be calculated also for a randomly or chaotically
time-dependent field $\mathbf{W}(\mathbf{r},t)$.

\begin{figure}[t]
\centering
\includegraphics[width=0.28\textwidth]{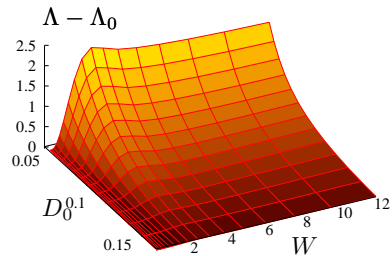}
\caption{(Color online) Lyapunov exponent characterizing stability
of the global mode mixed by periodically blinking vortex
\reff{eq20}. A similar picture holds for a stationary vortex.}
\label{fig3}\vspace{-2mm}
\end{figure}

We have calculated the LE  for a linearized two-dimensional
reaction-advection-diffusion equation \reff{eq801} subject to a
constant  open flow with a superimposed oscillating vortex,
described by the stream function
\be
\Psi(x,y)=V y+W\exp[-(x^2+y^2)R^{-2}]\cos(\omega t)\;.
\ee{eq20}
We fixed $V=1$, $R=1$ and $\omega=2$, and calculated the LE
$\Lambda$ for different molecular diffusion constants $D_0$ and
vortex intensities $W$ (Fig.~\ref{fig3}). Note that the parameter
$a$ simply shifts the LE, therefore we plot $\Lambda-\Lambda_0$,
% inFig.~\ref{fig3},
where $\Lambda_0=a-V^2/(4D_0)$ is the LE for a
non-mixed flow.
% with $W=0$.
 One can see that the mixing-induced
enhancement of field growth is mostly pronounced for small
diffusion and is maximal at $W\approx 3$. This is the mixing
strength at which a chaotic saddle~\cite{Tel_etal-05} in the
Lagrangean particle trajectories appears. A further increase of
the vortex intensity does not lead, however, to a significant
growth of the LE.
%The solution to the linearized problem (see
%Fig.~\ref{fig3}) defines the critical parameters (e.g., the mixing
%intensity), at which the global mode appears.

Now we develop a nonlinear theory of the global mode. It is clear
that the nonlinear saturation stops the exponential growth of a
slightly supercritical linear mode and leads to a nonlinear
solution with a finite amplitude. Our aim is to describe the
dependence of this amplitude on the deviation from criticality.
First we notice, that the very notion of the amplitude is here
nontrivial. Indeed, the nonlinear solution looks as in
Fig.~\ref{fig1}b (cf. Fig.~\ref{fig5} below); it saturates to
$u=1$ in the downstream direction. However, outside the mixing
region the field looks like a solution caused by a localized field
source. Thus, we can take the effective intensity of this source
$\e_{eff}$, which is proportional to the characteristic field
amplitude in the mixing region $u(0)$ [see relations (\ref{eq01}),
(\ref{eq02})], as the order parameter of the transition. The
deviation from the criticality we will measure with the growth
rate $\lambda$, for which holds $\lambda\propto W-W_{cr}$ in the
full model \reff{eq1} or $\lambda\propto D-D_{cr}$ in model
\reff{eq3}.

We will consider the simplest possible setup, namely the nonlinear
modification of one-dimensional Eq.~\reff{eq4}:
\be \frac{\partial u}{\partial t}+V\frac{\partial u}{\partial
x}=\frac{\partial}{\partial x}\left[ D(x)\frac{\partial
u}{\partial x}\right]+au(1-u^p)\;. \ee{eq81}
We look for a stationary global mode $u(x)$, and rewrite this equation as the system
\vspace{-2mm}
\begin{align}
\frac{dz}{dx}&=-z^2+\frac{V-\frac{d D(x)}{dx}}{D}z-\frac{a}{D}+\frac{a}{D}u^{p}\;,\label{eq9-1}\\
\frac{du}{dx}&=zu\;.\label{eq9-2}
\end{align}
\vspace{-3mm}

We consider this system separately in two spatial domains. The
first, linear region, includes the inflow and the mixing domains
(``source'' in Fig.~\ref{fig1}): $-\infty<x<x_r$, where the field
$u(x)$ remains small. In the second, outflow region
$x_r<x<\infty$, the field $u$ further grows (``tail'') and
nonlinearly saturates (``plateau''). In the linear region, because
of smallness of the field, we can neglect $u^p$ in \reff{eq9-1},
thus we obtain an equation similar to \reff{eq5}. The only
difference is that because we look for a stationary solution, in
\reff{eq9-1} the term $\sim \lambda$ is absent. Near the
criticality, where $\lambda$ is small, we can consider this term
as a perturbation, therefore the solution of \reff{eq9-1} in the
linear region is close to the solution of Eq.~\reff{eq5}; it has
the asymptotic $z\to \mu_+$ as $x\to -\infty$. Due to the
perturbation term $\propto \lambda$, at the right border of the
linear region $z$ deviates from $\mu_-$: the deviation $\mu_{-}-
z(x_r)$ is proportional to $\lambda$, and, thus, to $D-D_{cr}$. At
$x_r$ the field $u$ is small and $u(x_r)\propto u(0)$.

Next we consider full equations \reff{eq9-1}, \reff{eq9-2} in
the nonlinear region $x>x_r$. Here the solution should tend at
$x\to\infty$ to the saddle fixed point $u=1$, $z=0$. Thus,
starting integration from large values of $x$ in the negative
direction, we have to follow the stable manifold of this saddle
and match this solution at $x=x_r$ with the obtained above.
Because the value to be matched $z(x_r)$ is very close $\mu_-$, in
the region where the solution $(z,u)$ approaches $(z(x_r),u(x_r))$
we can write $z^2\approx\mu_-^2-2\mu_-\Delta z$ to obtain
\be \frac{d}{dx}\Delta z=(\mu_+-\mu_-)\Delta
z-\frac{a}{D}u^p(x_r)e^{p\mu_- (x-x_r)}\;. \ee{eq10}
Here, since $\Delta z=\mu_--z(x)$ is small, we have approximated
the solution of \reff{eq9-2} as $u\approx u(x_r)e^{\mu_-
(x-x_r)}$. Because linear inhomogeneous Eq.~\reff{eq10} is solved
in the \textit{negative} in $x$ direction, the solution follows
the \textit{slowest} exponent: $\Delta z\propto \exp[\gamma
(x-x_r)]$, where $\gamma=\text{min}(\mu_+-\mu_-,p\mu_-)$.

At the criticality, the region of validity of the exponential
solution $\Delta z\propto \exp[\gamma (x-x_r)]$ becomes very
large. Thus it is dominant for small deviations from criticality
$\lambda$, therefore we can estimate the coordinate $x_s$  at
which the field $u$ saturates (i.e., we reach the state $u\approx
1$ and $z\approx 0$) from the relations above: from $-\mu_-\approx
(z(x_r)-\mu_-)e^{\gamma (x_s-x_r)}$ it follows $(x_s-x_r) \approx
-\gamma^{-1}\ln (\mu_- - z(x_r))$. Substituting this in the
expression for $u(x)$, we obtain $u(x_r)\propto(\mu_- -
z(x_r))^{\mu_- /\gamma}$. Now we take into account that $\mu_- -
z(x_r)\propto D-D_{cr}$, and,  because the evolution of $u$ in the
interval $0<x<x_r$ only weakly depends on the criticality,
$\e_{eff}= u(0)\propto u(x_r)$. The final expression for the
scaling law of the amplitude of the global mode thus reads
\begin{gather}
\e_{eff}  \propto \lambda^\beta \;, %\propto (D-D_{cr})^\beta
% \label{eq120}
 \qquad \beta=\frac{\mu_-}{\gamma}=\text{max}\left(\frac{\mu_-}{\mu_+-\mu_-},\frac{1}{p}\right)\;.\label{eq12}
\end{gather}
The critical index $\beta$ depends only on the nonlinearity index $p$ and on the dimensionless velocity $v=V/V_f$:
\begin{equation}
\begin{aligned}
\beta
%=&\text{max}\left(\frac{v-\sqrt{v^2-1}}{2\sqrt{v^2-1}},\frac{1}{p}\right)
%\nonumber
%\\
=&
\begin{cases}
p^{-1}&\text{ if } v>\frac{2+p}{2\sqrt{1+p}}\;,\\
\frac{v-\sqrt{v^2-1}}{2\sqrt{v^2-1}}&\text{ if } 1<v<\frac{2+p}{2\sqrt{1+p}}\;.
\end{cases}
\end{aligned}
\label{eq13}
\end{equation}

\begin{figure}[tb]
\centering
\includegraphics[width=0.32\textwidth]{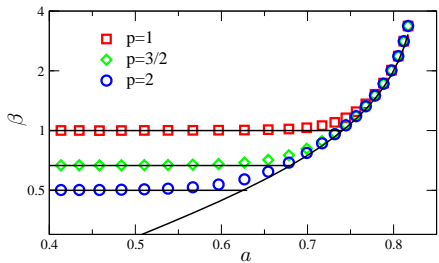}
\caption{(Color online) The critical exponent $\beta$ calculated
for the model \reff{eq1}  (symbols), compared with theoretical
prediction \reff{eq13} (lines). One can clearly see the crossover
between two regimes of the field saturation in dependence on
parameter $a$, the latter is related to $v$ in \reff{eq13}  via
$v\sim a^{-1/2}$.} \vspace{-2mm} \label{fig4}
\end{figure}

This main result of our letter can be physically interpreted as
follows. The exponent $\beta$ is determined solely by the
nonlinearity index $p$ if the throughflow velocity is much larger
than the front velocity ($v$ large). Here the field in the plateau
domain (see Fig.~\ref{fig1}) is effectively uncoupled from the
source, and the saturation of the instability is due to the local
nonlinearity at the source. For a small throughflow velocity ($v$
close to one) the plateau state interacts with the source via the
tail. Due to this ``remote control,'' the field at the source is
saturated more efficiently than due to nonlinearity, here the
exponent $\beta$ is determined solely by the form of the tail,
which depends on the velocities ratio $v$.

Below we check formula \reff{eq13} with direct numerical
simulations of model \reff{eq1}.  A stationary vortex \reff{eq20} with $\omega=0$ and $R=1$ was imposed on a constant
flow with $V=1$. Keeping the diffusion constant fixed $D_0=0.3$, for
different field growth rates $a$ we have found, from the
linearized equations, the critical vortex intensities $W_{cr}$ at
which the global mode becomes first unstable. Then we solved full
nonlinear equations close to criticality and found the exponent
$\beta$ according to \reff{eq12}. The stationary problem was
solved with a finite difference method in a domain $0\leq x\leq
60$, $0\leq y\leq 40$ with periodic boundary conditions in $y$ and
conditions $u(0)=0$, $\frac{\partial u}{\partial x}(60)=0$. The
results are presented in Fig.~\ref{fig4}, they are in a very good
agreement with the theoretical prediction \reff{eq12},
\reff{eq13}. Figure~\ref{fig5} shows the example of the stationary
mode appearing beyond the instability threshold.

\begin{figure}[!t]
%\vspace{5mm}
\centering
\includegraphics[width=0.35\textwidth]{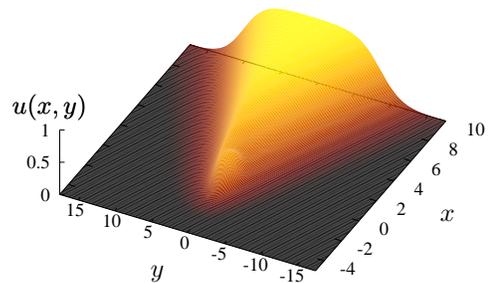}
\caption{(Color online) The active field behind a vortex with $W=4$ placed at $x=y=0$, for $V=1$, $a=0.5$, $D_0=0.3$, $p=2$. }
\label{fig5}
\end{figure}

In summary, we have described the mixing-induced transition from a
convectively unstable active field in an open flow to a persistent
global mode. Our theoretical approach bases on the representation
of the mixing region as a domain with locally enhanced effective
diffusion. For a nonlinear regime close to criticality we have
derived the critical exponent $\beta$ \reff{eq13} that depends
only on two parameters of the system: the dimensionless flow
velocity $v$ normalized by that of the front, and the nonlinearity
index $p$. For large velocities the critical exponent depends only
on the system's nonlinearity, which means a local in space
saturation of the instability. For small velocities the exponent
is a function of velocity, here the growing downstream tail of the
active field imposes the saturation. Notably, this prediction of
the one-dimensional theory is in a good accordance with
two-dimensional calculations.

We thank E. Knobloch, U. Feudel, S. Kuznetsov, N. Shnerb, and B.
Eckhardt for discussions and DFG
(SPP 1164 ``Nano- and Microfluidics'') for support.
%
%%%%%%%%%%%%%%%%%%%%%%%%%  references  %%%%%%%%%%%%%%%%%%%%%%%%%%
%%\bibliographystyle{apsrev}
%%\bibliography{nld-old,nld-current,%
%%pap-ab,pap-ce,pap-fg,pap-hj,pap-kl,pap-mn,pap-oq,pap-rs,pap-tz,%
%%pik,books,n-stand}

\end{document}